\documentclass{article}
\usepackage{spconf,amsmath,graphicx,hyperref, cite}
\usepackage{booktabs}
\usepackage{tabularx}
\usepackage{tikz}
\usepackage{enumitem}
\usepackage{xcolor}
\usepackage{microtype}

\usepackage{cite}
\let\oldbibliography\thebibliography
\renewcommand{\thebibliography}[1]{\oldbibliography{#1}
                                   \setlength{\itemsep}{-0.25mm}
                                   \vspace*{-0mm}}

\usetikzlibrary{shapes.geometric, arrows, calc}
\tikzstyle{block} = [rectangle, rounded corners, minimum width=3.5cm, minimum height=3.5cm, text centered, draw=black, fill=gray!10]
\tikzstyle{kmeans-node} = [circle, draw=black, fill=white, minimum size=4pt, inner sep=0pt]
\tikzstyle{kmeans-cluster} = [circle, draw=black, fill=black, minimum size=5pt, inner sep=0pt]
\tikzstyle{empty} = []
\tikzstyle{arrow} = [thick,->,>=to]
\tikzstyle{kmeans-arrow} = [thin, ->]

\newcolumntype{C}{>{\centering\arraybackslash}X}
\newcolumntype{L}{>{\raggedright\arraybackslash}X}
\newcolumntype{R}{>{\raggedleft\arraybackslash}X}
\newcolumntype{P}[1]{>{\raggedright\arraybackslash}p{#1}}

\newcommand{\tablesep}{\vspace{3pt}}

\title{
Unsupervised Lexicon Learning from Speech is Limited by Representations Rather Than Clustering
}

\name{Danel Slabbert, Simon Malan, Herman Kamper} 
\address{Electrical and Electronic Engineering, 
Stellenbosch University, South Africa}

\usepackage[prependcaption,textsize=scriptsize]{todonotes}
\definecolor{mycolor}{HTML}{FF6600}
\definecolor{coolcolor}{HTML}{27AE60}

\newcommand{\danel}[1]{\textcolor{coolcolor}{#1}}

\begin{document}
\ninept
\maketitle

\begin{abstract}
Zero-resource word segmentation and clustering systems aim to tokenise speech into word-like units without access to text labels.
Despite progress, the induced lexicons are still far from perfect.
In an idealised setting with gold word boundaries, we ask whether performance is limited by the representation of word segments, or by the clustering methods that group them into word-like types.
We combine a range of self-supervised speech features (continuous/discrete, frame/word-level) with different clustering methods ($k$-means, hierarchical, graph-based) on English and Mandarin data.
The best system uses graph clustering with dynamic time warping on continuous features.
Faster alternatives use graph clustering with cosine distance on averaged continuous features or edit distance on discrete unit sequences.
Through controlled experiments that isolate either the representations or the clustering method, we demonstrate that representation variability across segments of the same word type---rather than clustering---is the primary factor limiting performance. 
\end{abstract}


\section{Introduction}

Unsupervised word discovery aims to discover word-like units from unlabelled speech by locating word boundaries and clustering the resulting segments into hypothesised lexical categories~\cite{Ludusan2014}.
This is difficult because, unlike text, speech is continuous and lacks explicit word delimiters~\cite{Rsnen2012}. 
Moreover, different instances of the same word exhibit substantial acoustic variability, even when produced by the same speaker. 
Despite these challenges, human infants can discriminate between words in their native language within their first year \cite{Saffran1996, Jusczyk2002, Bergelson2012}.
Developing computational models that mimic this ability could therefore both inform theories of language acquisition \cite{Dupoux2018} and support the development of low-resource speech technology \cite{Besacier2014}.

Early approaches to unsupervised word discovery sought to identify recurring patterns across speech utterances, often relying on dynamic time warping (DTW)~\cite{Park2008}.
Despite advances~\cite{Rsnen2020, vanNiekerk2024}, these techniques still struggle to segment the entire input into meaningful units.
To address this, full-coverage methods have been proposed which aim to tokenise all of the input speech into word-like units~\cite{Lee2015,Rsnen2015, Kamper2017es,Bhati2020,Bhati2021,Algayres2022,Cuervo2022,Okuda2023,Kamper2023,Malan2024}.

Full-coverage systems can typically be broken into three components: an unsupervised word boundary detection component, a feature extraction method for representing word-like segments, and a clustering method used to group the hypothesised segments into a lexicon of word-like types. 
To assess the influence of boundary detection, Malan et al.~\cite{Malan2025} performed word discovery using ground-truth word boundaries.
Even in this idealised setting, the resulting lexicon is still far from perfect.
Since true boundaries were used, the errors stem either from the representation or clustering method.

This paper investigates these limitations by revisiting the task of unsupervised lexicon learning, where unlabelled word segments with true boundaries need to be clustered into hypothesised word types. 
We combine a range of representations from self-supervised learned (SSL) speech models with different clustering \danel{methods}.
We consider both discrete and continuous frame-level features. 
These are either kept in sequence form to represent word segments, with distances between sequences used for clustering, or averaged to obtain fixed-dimensional acoustic word embeddings~\cite{Levin2013, Sanabria2023}.
For clustering, we consider $k$-means, agglomerative, BIRCH, and graph clustering.
In experiments on English and Mandarin, our best system uses graph clustering with DTW over continuous sequences.
While producing a better lexicon than previous work~\cite{Malan2025}, the result remains imperfect, with word-level purity below 90\%.
This system is also much slower than those using averaged embeddings or discrete sequences. 

To definitively answer whether it is the representations or clustering that limits performance, we conduct two controlled experiments.
First, we reduce variability in clustering by forcing all instances of a word into the same cluster at initialisation.
Second, we eliminate representational inconsistency by artificially fixing all instances of the same word type to near-identical representations.
This allows the influence of speech representations and clustering methods to be assessed individually.
The results show that, despite substantial improvements, current speech representations, not clustering methods, constrain lexicon quality.
Thus, further research on SSL features is needed to make progress on unsupervised word discovery.

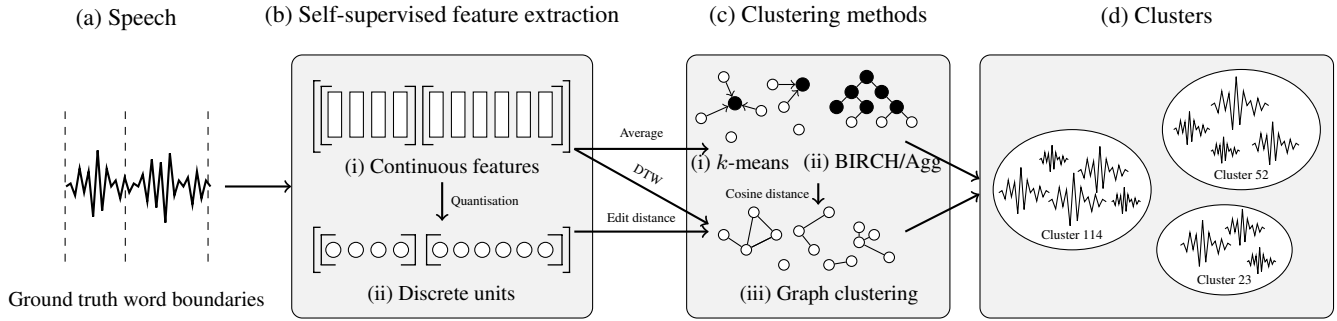
\begin{figure*}[t!]
  \begin{tikzpicture}[node distance=4cm]
    \node (n1) [empty, anchor=west] at (0,0) {};
    \node[anchor=south] at ($(n1.north) + (-1.2, 1.85)$) {(a) Speech};
    \begin{scope}[shift={([xshift=-2cm]n1.center)}, xscale=0.3, yscale=0.4]
    
         \draw[thick]
            (0,0) -- (0.2,0.1) -- (0.4,-0.3) -- (0.6,0.6) -- (0.8,-0.2)
            -- (1.0,0.5) -- (1.2,-1.0) -- (1.4,1.2) -- (1.6,-0.8)
            -- (1.8,0.6) -- (2.0,0.0) -- (2.2,0.2) -- (2.4,-0.4)
            -- (2.6,0.3) -- (2.8,-0.2) -- (3.0,0.1) -- (3.2,0.0)
            -- (3.4,0.4) -- (3.6,-0.5) -- (3.8,0.9) -- (4.0,-0.7)
            -- (4.2,0.8) -- (4.4,-1.2) -- (4.6,1.1) -- (4.8,-0.3)
            -- (5.0,0.2) -- (5.2,0.0) -- (5.4,0.3) -- (5.6,-0.6)
            -- (5.8,0.5) -- (6.0,-0.4) -- (6.2,0.2) -- (6.4,0.0);
        
    \end{scope}
    \draw[dashed, thin] (0, -1) -- (0, 1);
    \draw[dashed, thin] (-1.1, -1) -- (-1.1, 1); 
    \draw[dashed, thin] (-1.9, -1) -- (-1.9, 1);
    \node[anchor=south, font=\footnotesize] at (-0.95, -1.7) {Ground truth word boundaries};

    \node (n2) [block, right of=n1,  minimum width=4cm, node distance=3cm] {};
    \node[anchor=south] at ($(n2.north) + (0, 0.25)$) {(b) Self-supervised feature extraction};
    \node[anchor=west] at ($(n2.north west)+(0.6,-1.5)$) {\footnotesize (i) Continuous features};
    \draw[line width=0.5pt]
        ($(n2.north west)+(0.3,-0.35)$) -- ++(0, -0.95)       
        ($(n2.north west)+(0.291,-0.35)$) -- ++(0.1, 0)        
        ($(n2.north west)+(0.291,-1.3)$) -- ++(0.1, 0);       
        
    \foreach \i in {1,...,4} {
        \draw[fill=white] ($(n2.north west)+(0.3,-1.1*\i*0.0 - 0.8)$) ++(\i*0.29-0.1,-0.3) 
            rectangle ++(0.18,0.6);
    }
   
    \draw[line width=0.5pt]
    ($(n2.north west)+(0.4,-0.425)$) -- ++(0, -0.775)       
    ($(n2.north west)+(0.391,-0.425)$) -- ++(0.15, 0)        
    ($(n2.north west)+(0.391,-1.2)$) -- ++(0.15, 0);       
    
    \draw[line width=0.5pt]
    ($(n2.north west)+(1.65,-0.425)$) -- ++(0, -0.775)       
    ($(n2.north west)+(1.658,-0.425)$) -- ++(-0.15, 0)       
    ($(n2.north west)+(1.658,-1.2)$) -- ++(-0.15, 0);      

    \foreach \i in {1,...,6} {
        \draw[fill=white] ($(n2.north west)+(1.65,-1.1*\i*0.0 - 0.8)$) ++(\i*0.29-0.1,-0.3) 
            rectangle ++(0.18,0.6);
    }
    \draw[line width=0.5pt]
    ($(n2.north west)+(1.75,-0.425)$) -- ++(0, -0.775)       
    ($(n2.north west)+(1.741,-0.425)$) -- ++(0.15, 0)        
    ($(n2.north west)+(1.741,-1.2)$) -- ++(0.15, 0);       
    
    \draw[line width=0.5pt]
    ($(n2.north west)+(3.6,-0.425)$) -- ++(0, -0.775)       
    ($(n2.north west)+(3.609,-0.425)$) -- ++(-0.15, 0)       
    ($(n2.north west)+(3.609,-1.2)$) -- ++(-0.15, 0);      

    \draw[line width=0.5pt]
        ($(n2.north west)+(3.7,-0.35)$) -- ++(0, -0.95)       
        ($(n2.north west)+(3.709,-0.35)$) -- ++(-0.1, 0)        
        ($(n2.north west)+(3.709,-1.3)$) -- ++(-0.1, 0);       

    \node[anchor=west] at ($(n2.north west)+(0.9,-3.2)$) {\footnotesize (ii) Discrete units};
    \draw[line width=0.5pt]
        ($(n2.north west)+(0.3,-2.3)$) -- ++(0, -0.6)       
        ($(n2.north west)+(0.291,-2.3)$) -- ++(0.1, 0)        
        ($(n2.north west)+(0.291,-2.9)$) -- ++(0.1, 0);       
    
    \foreach \i in {1,...,4} {
        \draw[fill=white] ($(n2.north west)+(0.38,-2.2)$) ++(\i*0.3-0.12,-0.4)
            circle (0.1);
    }

    \draw[line width=0.5pt]
        ($(n2.north west)+(0.4,-2.4)$) -- ++(0, -0.4)       
        ($(n2.north west)+(0.391,-2.4)$) -- ++(0.2, 0)        
        ($(n2.north west)+(0.391,-2.8)$) -- ++(0.2, 0);       

    \draw[line width=0.5pt]
        ($(n2.north west)+(1.65,-2.4)$) -- ++(0, -0.4)       
        ($(n2.north west)+(1.658,-2.4)$) -- ++(-0.2, 0)       
        ($(n2.north west)+(1.658,-2.8)$) -- ++(-0.2, 0);      

    \foreach \i in {1,...,6} {
        \draw[fill=white] ($(n2.north west)+(1.82,-2.2)$) ++(\i*0.28-0.12,-0.4)
            circle (0.1);
    }
    
    \draw[line width=0.5pt]
        ($(n2.north west)+(1.8,-2.4)$) -- ++(0, -0.4)       
        ($(n2.north west)+(1.791,-2.4)$) -- ++(0.2, 0)        
        ($(n2.north west)+(1.791,-2.8)$) -- ++(0.2, 0);       
    
    \draw[line width=0.5pt]
        ($(n2.north west)+(3.6,-2.4)$) -- ++(0, -0.4)       
        ($(n2.north west)+(3.609,-2.4)$) -- ++(-0.2, 0)       
        ($(n2.north west)+(3.609,-2.8)$) -- ++(-0.2, 0);      

    \draw[line width=0.5pt]
        ($(n2.north west)+(3.7,-2.3)$) -- ++(0, -0.6)       
        ($(n2.north west)+(3.709,-2.3)$) -- ++(-0.1, 0)       
        ($(n2.north west)+(3.709,-2.9)$) -- ++(-0.1, 0);      

    \node (n3) [block, right of=n2,  node distance=5cm] {};
    
    \node[anchor=south] at ($(n3.north) + (0, 0.25)$) {(c) Clustering methods};
    
    \begin{scope}[shift={( $(n3.north west) + (0.6 ,-1.1)$ )}]
        \node[anchor=west] at (-0.65,-0.35) {\footnotesize (i) $k$-means};
    
        \node(a)[kmeans-node] at  (0,0) {};
        \node(b)[kmeans-node] at  (-0.4,0.25) {};
        \node(c)[kmeans-node] at  (0.4, 0.35) {};
        \node(d)[kmeans-cluster] at  (0.05, 0.45) {};
        \node(e)[kmeans-node] at  (-0.1, 0.8) {};

        \node(f)[kmeans-node] at  (0.9, 0.1) {};
        \node(g)[kmeans-node] at  (0.55, 0.7) {};
        \node(h)[kmeans-cluster] at  (0.95, 0.7) {};
        \node(i)[kmeans-node] at  (0.7, 0.4) {};

        \draw[kmeans-arrow] (b) -- (d);
        \draw[kmeans-arrow] (c) -- (d);
        \draw[kmeans-arrow] (e) -- (d);       

        \draw[kmeans-arrow] (g) -- (h);
        \draw[kmeans-arrow] (i) -- (h);
        
    \end{scope}
    \begin{scope}[shift={( $(n3.north west) + (1.8 ,-1.1)$ )}]
        \node[anchor=west] at (-0.35,-0.35) {\footnotesize (ii) BIRCH/Agg};
        \node(a)[kmeans-cluster] at  (0.6, 0.8) {};
        \node(b)[kmeans-cluster] at  (0.4, 0.6) {};
        \node(c)[kmeans-cluster] at  (0.2, 0.4) {};
        \node(d)[kmeans-cluster] at  (0.6, 0.4) {};
        \node(e)[kmeans-node] at  (0.4, 0.2) {};
        
        \node(f)[kmeans-cluster] at  (0.8, 0.6) {};
        \node(g)[kmeans-cluster] at  (1, 0.4) {};
        \node(h)[kmeans-node] at  (0.8, 0.2) {};
        \node(i)[kmeans-node] at  (1.2, 0.2) {};

        \draw[] (a) -- (b);
        \draw[] (b) -- (c);
        \draw[] (b) -- (d);
        \draw[] (d) -- (e);

        \draw[] (a) -- (f);
        \draw[] (f) -- (g);
        \draw[] (g) -- (h);
        \draw[] (g) -- (i);

    \end{scope}
    
    \begin{scope}[shift={( $(n3.north west)+(1.3,-2.9)$ )}]
        \node[anchor=west] at (-0.7,-0.3) {\footnotesize (iii) Graph clustering};

        \node(a)[kmeans-node] at  (0, 0.1) {};
        \node(b)[kmeans-node] at  (-0.1, 0.5) {};
        \node(c)[kmeans-node] at  (-0.5, 0.3) {};
        \node(d)[kmeans-node] at  (-0.8, 0.5) {};
        \node(e)[kmeans-node] at  (-0.4, 0.8) {};
        
        \node(f)[kmeans-node] at  (0.2, 0.55) {};
        \node(g)[kmeans-node] at  (0.4, 0.3) {};
        \node(h)[kmeans-node] at  (0.6, 0.1) {};
        \node(i)[kmeans-node] at  (0.9, 0.15) {};
        \node(j)[kmeans-node] at  (1.2, 0.5) {};
        \node(k)[kmeans-node] at  (0.6, 0.8) {};
        \node(l)[kmeans-node] at  (1, 0.6) {};
        \node(m)[kmeans-node] at  (1, 0.4) {};
        \node(n)[kmeans-node] at  (1.4, 0.2) {};

        \draw[] (b) -- (c);    
        \draw[] (e) -- (c);
        \draw[] (e) -- (b);
        \draw[] (b) -- (c);   
        \draw[] (c) -- (d);    

        \draw[] (g) -- (f);    
        \draw[] (f) -- (k);

        \draw[] (j) -- (m);
        \draw[] (m) -- (l);
        \draw[] (m) -- (n);    

        \draw[] (h) -- (i);

    \end{scope}
    \node (n4) [block, right of=n3, minimum width=4.7cm, node distance=4.5cm] {};
    \node[anchor=south] at ($(n4.north) + (0, 0.28)$) {(d) Clusters};
    
      \begin{scope}[shift={(n4.north)}, yshift=-1cm, xscale=0.3, yscale=0.4]
        \draw[fill=white] (-3.75, -2) ellipse (3.5 and 2);
        \node at (-3.75, -3.5) {\tiny Cluster 114};
        
        \begin{scope}[scale=0.7, shift={(-5, -2)}]
             \draw[]
                (0,0) -- (0.2,0.1) -- (0.4,-0.3) -- (0.6,0.6) -- (0.8,-0.2)
                -- (1.0,0.5) -- (1.2,-1.0) -- (1.4,1.2) -- (1.6,-0.8)
                -- (1.8,0.6) -- (2.0,0.0) -- (2.2,0.2) -- (2.4,-0.4)
                -- (2.6,0.3) -- (2.8,-0.2) -- (3.0,0.1) -- (3.2,0.0);
        \end{scope}
        \begin{scope}[scale=0.7, shift={(-10, -3)}]
          \draw[]
                (0,0) -- (0.2,0.1) -- (0.4,-0.3) -- (0.6,0.6) -- (0.8,-0.2)
                -- (1.0,0.5) -- (1.2,-1.0) -- (1.4,1.2) -- (1.6,-0.8)
                -- (1.8,0.6) -- (2.0,0.0) -- (2.2,0.2) -- (2.4,-0.4)
                -- (2.6,0.3) -- (2.8,-0.2) -- (3.0,0.1) -- (3.2,0.0);
        \end{scope}
        \begin{scope}[scale=0.4, shift={(-13, -2.5)}]
            \draw[]
                (0,0) -- (0.2,0.1) -- (0.4,-0.3) -- (0.6,0.6) -- (0.8,-0.2)
                -- (1.0,0.5) -- (1.2,-1.0) -- (1.4,1.2) -- (1.6,-0.8)
                -- (1.8,0.6) -- (2.0,0.0) -- (2.2,0.2) -- (2.4,-0.4)
                -- (2.6,0.3) -- (2.8,-0.2) -- (3.0,0.1) -- (3.2,0.0);
        \end{scope}
        \begin{scope}[scale=0.8, shift={(-6, -3)}]
            \draw[]
                (0,0) -- (0.2,0.1) -- (0.4,-0.3) -- (0.6,0.6) -- (0.8,-0.2)
                -- (1.0,0.5) -- (1.2,-1.0) -- (1.4,1.2) -- (1.6,-0.8)
                -- (1.8,0.6) -- (2.0,0.0) -- (2.2,0.2) -- (2.4,-0.4)
                -- (2.6,0.3) -- (2.8,-0.2) -- (3.0,0.1) -- (3.2,0.0);           
        \end{scope}
        \begin{scope}[scale=0.4, shift={(-5, -6)}]
            \draw[]
                (0,0) -- (0.2,0.1) -- (0.4,-0.3) -- (0.6,0.6) -- (0.8,-0.2)
                -- (1.0,0.5) -- (1.2,-1.0) -- (1.4,1.2) -- (1.6,-0.8)
                -- (1.8,0.6) -- (2.0,0.0) -- (2.2,0.2) -- (2.4,-0.4)
                -- (2.6,0.3) -- (2.8,-0.2) -- (3.0,0.1) -- (3.2,0.0);
        \end{scope}

        \draw[fill=white] (3, -4) ellipse (3 and 1.5); 
        \node at (3, -5) {\tiny Cluster 23};
        \begin{scope}[scale=0.7, shift={(1.5, -5.5)}]
            \draw[]
                (0,0) -- (0.2,0.1) -- (0.4,-0.3) -- (0.6,0.6) -- (0.8,-0.2)
                -- (1.0,0.5) -- (1.2,-1.0) -- (1.4,1.2) -- (1.6,-0.8)
                -- (1.8,0.6) -- (2.0,0.0) -- (2.2,0.2) -- (2.4,-0.4)
                -- (2.6,0.3) -- (2.8,-0.2) -- (3.0,0.1) -- (3.2,0.0);
        \end{scope}
        \begin{scope}[scale=0.4, shift={(10, -11)}]
          \draw[]
                (0,0) -- (0.2,0.1) -- (0.4,-0.3) -- (0.6,0.6) -- (0.8,-0.2)
                -- (1.0,0.5) -- (1.2,-1.0) -- (1.4,1.2) -- (1.6,-0.8)
                -- (1.8,0.6) -- (2.0,0.0) -- (2.2,0.2) -- (2.4,-0.4)
                -- (2.6,0.3) -- (2.8,-0.2) -- (3.0,0.1) -- (3.2,0.0);
        \end{scope}
        \begin{scope}[scale=0.55, shift={(5.5, -6)}]
            \draw[]
                (0,0) -- (0.2,0.1) -- (0.4,-0.3) -- (0.6,0.6) -- (0.8,-0.2)
                -- (1.0,0.5) -- (1.2,-1.0) -- (1.4,1.2) -- (1.6,-0.8)
                -- (1.8,0.6) -- (2.0,0.0) -- (2.2,0.2) -- (2.4,-0.4)
                -- (2.6,0.3) -- (2.8,-0.2) -- (3.0,0.1) -- (3.2,0.0);
                      
        \end{scope}

        \draw[fill=white] (3.75, 0) ellipse (3.5 and 2); 
        \node at (3.75, -1.5) {\tiny Cluster 52};
        \begin{scope}[scale=0.5, shift={(1.5, 0)}]
            \draw[]
                (0,0) -- (0.2,0.1) -- (0.4,-0.3) -- (0.6,0.6) -- (0.8,-0.2)
                -- (1.0,0.5) -- (1.2,-1.0) -- (1.4,1.2) -- (1.6,-0.8)
                -- (1.8,0.6) -- (2.0,0.0) -- (2.2,0.2) -- (2.4,-0.4)
                -- (2.6,0.3) -- (2.8,-0.2) -- (3.0,0.1) -- (3.2,0.0);
        \end{scope}
        \begin{scope}[scale=0.8, shift={(3, 1)}]
          \draw[]
                (0,0) -- (0.2,0.1) -- (0.4,-0.3) -- (0.6,0.6) -- (0.8,-0.2)
                -- (1.0,0.5) -- (1.2,-1.0) -- (1.4,1.2) -- (1.6,-0.8)
                -- (1.8,0.6) -- (2.0,0.0) -- (2.2,0.2) -- (2.4,-0.4)
                -- (2.6,0.3) -- (2.8,-0.2) -- (3.0,0.1) -- (3.2,0.0);
        \end{scope}
        \begin{scope}[scale=0.4, shift={(6, -1.8)}]
           \draw[]
                (0,0) -- (0.2,0.1) -- (0.4,-0.3) -- (0.6,0.6) -- (0.8,-0.2)
                -- (1.0,0.5) -- (1.2,-1.0) -- (1.4,1.2) -- (1.6,-0.8)
                -- (1.8,0.6) -- (2.0,0.0) -- (2.2,0.2) -- (2.4,-0.4)
                -- (2.6,0.3) -- (2.8,-0.2) -- (3.0,0.1) -- (3.2,0.0);
        \end{scope}
        \begin{scope}[scale=0.65, shift={(6.5, -0.8)}]
            \draw[]
                (0,0) -- (0.2,0.1) -- (0.4,-0.3) -- (0.6,0.6) -- (0.8,-0.2)
                -- (1.0,0.5) -- (1.2,-1.0) -- (1.4,1.2) -- (1.6,-0.8)
                -- (1.8,0.6) -- (2.0,0.0) -- (2.2,0.2) -- (2.4,-0.4)
                -- (2.6,0.3) -- (2.8,-0.2) -- (3.0,0.1) -- (3.2,0.0);
                      
        \end{scope}
    \end{scope}
    \draw [arrow] (n1) -- (n2);
    \draw [arrow] ($(n2.east) + (-0.25, 0.5)$) -- node[above]{\tiny Average} ($(n3.west) + (0.3, 0.5)$);
    \draw [arrow] ($(n2.north) + (0, -1.7)$) -- node[right]{\tiny Quantisation} ($(n2.south) + (0, +1.3)$);
    \draw [arrow] ($(n2.east) + (-0.25, 0.5)$) --  node[sloped, above]{\tiny DTW} 
 ($(n3.west) + (0.3, -0.5)$);
    \draw [arrow] ($(n2.east) + (-0.25, -0.6)$) -- node[above]{\tiny Edit distance} ($(n3.west) + (0.3, -0.6)$);
    \draw [arrow] ($(n3.south) + (0.0, +1.8)$) -- node[left]{\tiny Cosine distance} ($(n3.north) + (0, -2)$);
    \draw [arrow] ($ (n3.east) + (-0.6, 0.6)$) -- ($ (n4.west) + (0, 0.1)$);
    \draw [arrow] ($ (n3.east) + (-0.6, -0.6)$) -- ($ (n4.west) + (0, -0.1)$);
    \end{tikzpicture}

  \caption{We consider different systems of (b) speech representations and (c) clustering approaches in order to learn (d) a lexicon from (a)~unlabelled speech. 
  For this study, we assume that true word boundaries are known.}
  \label{fig:method_figure}
\end{figure*}

\section{Methods}
\label{sec:methods}

We investigate several combinations of speech representation and clustering methods to understand the current landscape of unsupervised lexicon learning.
The systems are shown in Figure~\ref{fig:method_figure}.
Apart from~\cite{Malan2025}, who implicitly looked at lexicon learning, work on this task is limited. One exception is~\cite{Kamper2014}, which applied different clustering approaches on acoustic word embeddings. But this was done using conventional features in a time before neural representations.

\subsection{Representations}

Given ground truth word segments (Figure~\ref{fig:method_figure}-a), we start by extracting continuous features from intermediate layers of SSL speech models (Figure~\ref{fig:method_figure}-b-i).
We consider a range of SSL models, all trained using a masked prediction pretext task~\cite{Hsu2021, Chen2022, vanNiekerk2022, Boito2024}.
The different SSL models are described in detail in Section~\ref{sec:exp_setup_implementation}.
As an alternative to continuous features, discretised representations are obtained by quantising continuous features, producing a lower-bitrate encoding which reduces downstream computational load (Figure~\ref{fig:method_figure}-b-ii).
Previous work~\cite{vanNiekerk2024} showed that quantisation can remove speaker-specific information from continuous features, potentially improving consistency in representations used for lexicon learning.

While some clustering approaches can be applied directly to feature sequences, others require word segments to be represented as single fixed-dimensional vectors.
A common approach is to use an acoustic word embedding obtained 
by averaging the features of a word segment across the temporal dimension~\cite{Levin2013}.
While simple, several studies have shown that averaging appropriate SSL features results in 
robust acoustic word embeddings~\cite{Sanabria2023, Jacobs2024, Herreilers2025}.

\subsection{Clustering Methods} 

We couple the different representations with a range of clustering methods (Figure~\ref{fig:method_figure}-c), resulting in six lexicon learning systems.
The first four systems cluster averaged acoustic word embeddings using one of the following methods (Figure~\ref{fig:method_figure}-c): $k$-means clustering; BIRCH clustering~\cite{Zhang1996}, an efficient tree-based algorithm; agglomerative hierarchical clustering~\cite{Nielsen2016}; or graph clustering, where word segments are represented as nodes and edges are weighted by the cosine similarity between their embeddings.

The two remaining systems also use graph clustering, but operate on feature sequences, retaining temporal information.
The distance metric used for graph construction corresponds to the type of representations.
For continuous features, pairwise distances are computed using DTW, which aligns sequences of varying lengths and calculates the cost of the optimal alignment (Figure~\ref{fig:method_figure}-b-i to Figure \ref{fig:method_figure}-c-iii).
For discrete unit sequences, edit distance is used, which measures the minimum number of operations required to convert one sequence into another (Figure~\ref{fig:method_figure}-b-ii to Figure~\ref{fig:method_figure}-c-iii).
For all graph clustering systems, a graph is incrementally constructed with edges inserted between nodes when the similarity exceeds a predefined threshold.
The graph is then partitioned using the efficient Leiden algorithm \cite{Traag2019} with the constant Potts model \cite{Reichardt2004} as clustering objective. 

\section{Experimental Setup}
\label{sec:exp_setup}

\subsection{Data and Evaluation}
We use LibriSpeech \cite{Panayotov2015} dev-clean for development and test-clean for final evaluations. 
Each set contains 5.4 hours of English speech from 40 speakers. 
Word boundaries are obtained from forced alignments~\cite{McAuliffe2017}.
For the Mandarin evaluation, we use data from Track~2 of the ZeroSpeech challenge~\cite{Dunbar2017}, consisting of 2.5 hours of speech from 12 speakers with word alignments.
For the Mandarin experiments, the optimal hyperparameters from the English development experiments are retained as is.

We evaluate lexicon quality using a range of standard metrics.
Normalised edit distance~(NED)~\cite{Ludusan2014} measures the phonetic purity of clusters by calculating the length-normalised edit distance between phonetic transcriptions (from forced alignments) of all word segment pairs within a cluster.
A single average over all clusters gives the final NED; lower is better.
Purity measures how homogeneous each cluster is with respect to the true word labels.
For every cluster, the most frequent word type is identified, and the number of segments belonging to that type is counted. 
These are summed over all clusters and divided by the total number of segments in~the~dataset.
V-measure~\cite{Rosenberg2007} is the harmonic mean of homogeneity (a purity-like metric) and completeness (measuring the degree to~which segments of the same type are assigned to the same cluster).
Higher purity and V-measure are better.
Bitrate is the average data rate of~the encoded output, measured in bits per second (bits/s); lower is better. 

Purity and NED can be artificially improved by increasing the number of clusters.
To enable a fair comparison between methods, and since we are operating in an idealised setting, we fix the number of clusters to the true number of word types in each dataset.
The number of clusters for LibriSpeech dev-clean and test-clean is 8,216 and 8,006, respectively, while for Mandarin it is 8,871.
Graph clustering does not take the number of clusters as an explicit input but instead infers it from the data; we tune the hyperparameters to get approximately the desired number of clusters.

\subsection{Implementation: Representations and Clustering Methods}
\label{sec:exp_setup_implementation}

We compare speech representations from a range of SSL models, all producing 20 ms frames.
To test state-of-the-art features~\cite{Pasad2023, Pasad2024}, we use HuBERT Large~\cite{Hsu2021} and WavLM Large~\cite{Chen2022}.
A HuBERT Base model fine-tuned for voice-conversion, HuBERT Soft~\cite{vanNiekerk2022}, is also considered for its ability to preserve phonetic content while suppressing speaker-specific information: its continuous features are extracted from a linear projection layer that predicts a distribution over discrete units, thereby providing a middle-ground between continuous and discrete features.
To assess the effects of pre-training language, we use the multilingual HuBERT (mHuBERT)~\cite{Boito2024} pre-trained on 90k hours across 147 languages (including English and Mandarin).
For the Mandarin tests, we include a HuBERT Large model pre-trained on 10k hours of Mandarin.\footnote{\label{mandarin-hubert}https://huggingface.co/TencentGameMate/chinese-hubert-large}
For all large SSL models, 1,024-dimensional features are extracted from the 21st layer based on its ability to capture word information~\cite{Pasad2023,Pasad2024}.
Following~\cite{Malan2024}, 768-dimensional mHuBERT features are extracted from the eighth~layer.

All these features are subsequently mean and variance normalised, and then projected to 350 dimensions using principal component analysis (PCA).
This gave the best performance on dev-clean, and improved on the results from~\cite{Malan2024,Malan2025}, where 250 dimensions were used.
For HuBERT Soft, features from the 12th layer are pushed through the projection layer, resulting in 256-dimensional soft features, which are mean and variance normalised.
For each SSL model, acoustic word embeddings are obtained by averaging the different low-dimensional features and then normalising to the unit sphere.

Discrete features are obtained by applying $k$-means clustering with 500 clusters on the different SSL representations (without normalisation or PCA).
These unit extraction models are trained on 50 hours of audio from LibriSpeech train-clean for English data, or on the complete dataset for the Mandarin data.
Following \cite{Kamper2023, Visser2025}, we use duration-penalised dynamic programming to smooth the resulting sequences by discouraging rapid unit changes.
The unit sequences are not deduplicated, which worked better in development experiments.

We first describe the clustering methods that learn a lexicon on averaged acoustic word embeddings.
For $k$-means, we use the efficient FAISS library.\footnote{https://github.com/facebookresearch/faiss}
But unlike~\cite{Malan2025}, we switch the default random initialisation to $k$-means++ initialisation~\cite{Arthur2007}.
This selects the first centroid at random, with each subsequent centroid chosen from the remaining data points with probability proportional to the squared distance from the closest existing centroid.
This alternative initialisation gave large improvements over~\cite{Malan2024,Malan2025} on development data.
BIRCH and agglomerative clustering are implemented using scikit-learn.\footnote{https://scikit-learn.org}
We apply BIRCH with a distance threshold of 0.25.
Agglomerative clustering uses Ward linkage~\cite{Murtagh2014}, which merges clusters by attempting to minimise the total within-cluster variance. 

Graph clustering can be used with either averaged acoustic embeddings or feature sequences.
We use the efficient igraph library \cite{Csardi2006}.
Distance-based thresholds are applied globally: nodes with no edges 
remain isolated. 
Thresholds are tuned based on memory constraints and development performance, with values of 0.65 for edit distance graphs, 0.4 for cosine distance graphs, and 0.35 for DTW graphs.

\section{Experimental Results}
\subsection{A First Evaluation: Representations}

Before comparing the various combinations of representations and clustering methods, we start by selecting the best-performing SSL features.
Table~\ref{tab:feature_comparison} reports English development scores using two representative systems: continuous averaged features which are $k$-means clustered (top), and discrete unit sequences which are graph clustered using edit distance (bottom).

Systems using WavLM Large features perform well, especially in its continuous form (Table~\ref{tab:feature_comparison}-top) where it outperforms all other models.
In the discrete graph clustering system, HuBERT Large also performs well.
In contrast, mHuBERT generally results in worse lexicons, indicating that diluting English pre-training data with other languages worsens performance compared to using English-only models.
Although HuBERT Soft is competitive in the continuous setting, the discretised version (bottom) performs much worse.
Based on these results, we use WavLM Large features for the subsequent lexicon learning experiments on English data.

\begin{table}[t!]
    \centering
    \caption{
    Lexicon quality (\%) on LibriSpeech dev-clean when different SSL features are used in
    two representative systems, one using continuous and the other discrete features.
    }
    \tablesep
    \label{tab:feature_comparison}
    \renewcommand{\arraystretch}{1.1}
    \begin{tabularx}{\linewidth}{@{}l@{\ \ }RRr@{}}
    \toprule
    Features & NED & Purity & V-measure \\
    \midrule
    \multicolumn{4}{c}{Continuous + average + $k$-means} \\
    \midrule
    WavLM Large  & \textbf{7.4} & \textbf{89.3} & \textbf{83.7} \\
    HuBERT Large  & 9.3 & 89.0 & 83.6 \\
    HuBERT Soft & 10.0 & 85.0 & 83.1 \\
    mHuBERT & 10.8 & 83.4 & 82.2 \\
    \midrule
    \multicolumn{4}{c}{Discrete + edit distance + graph clustering} \\
    \midrule
    WavLM Large  & \textbf{7.3} & 83.3 & 88.6 \\
    HuBERT Large  & 7.8 & \textbf{85.0} & \textbf{89.8} \\
    HuBERT Soft & 23.5 & 59.6 & 78.9 \\
    mHuBERT & 29.7 & 61.0 & 79.1 \\
    \bottomrule
    \end{tabularx}
    \vspace{-3pt}
\end{table}

\subsection{Evaluation on English: Representations with Clustering}

Table \ref{tab:test_results} shows the lexicon learning performance on English test data of the different representation--clustering systems using WavLM Large features (continuous or discrete).
Continuous averaged features with graph clustering achieves the best purity (89.6\%), V-measure (90.3\%), and bitrate (35.6).
The continuous features with DTW and graph clustering yield the lowest NED (5.2\%), but at a substantial computational cost, being around 250 times slower than alternative systems.
These results all improve on the continuous averaged $k$-means approach from~\cite{Malan2024,Malan2025}, which achieves a NED of 17.3\%, purity of 78.2\%, V-measure of 80.0\%, and a bitrate of 41.4 (after altering the experimental setup to exactly match the protocol here).

\begin{table}[t]
    \centering
    \caption{
    Lexicon quality on LibriSpeech test-clean in terms of NED~(\%), purity  (\%), V-measure  (\%), bitrate (bits/second), and runtime (seconds) for six different representation--clustering systems.
    }
    \tablesep
    \label{tab:test_results}
        \renewcommand{\arraystretch}{1.1}
    \begin{tabularx}{\linewidth}{@{}l@{\ \ }RRRRr@{}}
    \toprule
    System & NED & Purity & V-m & Bitrate & Runtime \\
    \midrule
    continu + avg + $k$-means & 8.6 & 88.4 & 83.6 & 40.9 & 281.0 \\
    continu + avg + BIRCH & 6.8 & 89.5 & 84.1 & 41.0 & \textbf{415.0} \\
    continu + avg + agglom & 6.8 & 89.5 & 84.1 & 40.9 & 433.0 \\
    continu + avg + graph & 6.7 & \textbf{89.6} & \textbf{90.3} & \textbf{35.6} & 484.0 \\
    continu + DTW + graph & \textbf{5.2} & 89.3 & 89.1 & 36.6 & 123,630.9 \\
    discrete + edit + graph & 7.9 & 83.0 & 88.4 & 36.9 & 1,526.6 \\
    \bottomrule
    \end{tabularx} 
\end{table}

In general, continuous representations outperform discrete units on purity and NED, with the discrete graph system only being competitive in V-measure (88.4\%).
Of the different clustering methods, graph-based clustering consistently yields higher V-measure scores than simpler methods like $k$-means or agglomerative clustering.
Bitrate indicates that graph clustering produces a more compact encoding output, especially when used with the averaged embeddings.

Although all systems deliver competitive lexicons and improve on the results from~\cite{Malan2024,Malan2025}, none of them achieve a perfect lexicon, even when provided with a ground-truth word segmentation: NED shows that the clustered sequences still have phonetic differences, and none of the systems achieve perfect purity or completeness.
Before we investigate what limits performance, we see if this is also true on another language, Mandarin. 
Apart from being in a different language family, the SSL representations available in Mandarin are also different from those available in English.

\begin{table}[!t]
    \centering
    \caption{
    Lexicon quality (\%) on Mandarin data using different SSL features in a subset of relevant systems.
    }
    \tablesep
    \label{tab:mandarin_comparison}
        \renewcommand{\arraystretch}{1.1}

    \begin{tabularx}{\linewidth}{@{}l@{\ \ }RRr@{}}
        \toprule
        Features & NED & Purity & V-measure \\
        \midrule
        \multicolumn{4}{c}{Continuous + average + $k$-means} \\
        \midrule
        WavLM Large & 52.7 & 63.3 & 87.9 \\
        mHuBERT & 46.9 & 66.6 & 88.6 \\
        Mandarin HuBERT Large & \textbf{16.8} & \textbf{80.5} & \textbf{92.1} \\  
        \midrule
        \multicolumn{4}{c}{Continuous + average + cosine distance + graph clustering} \\
        \midrule
        WavLM Large & 43.4 & 64.3 & 88.5 \\
        mHuBERT & 39.1 & 67.5 & 89.3 \\
        Mandarin HuBERT Large & \textbf{4.9} & \textbf{82.8} & \textbf{94.2} \\
        \midrule
        \multicolumn{4}{c}{Continuous + DTW distance + graph clustering} \\
        \midrule
        WavLM Large & 33.2 & 68.3 & 89.6 \\
        mHuBERT & 30.3 & 71.5 & 90.5 \\
        Mandarin HuBERT Large & \textbf{5.8} & \textbf{82.0} & \textbf{93.6} \\
        \bottomrule
    \end{tabularx}
\end{table}

\subsection{Evaluation on Mandarin}

Table~\ref{tab:mandarin_comparison} shows how well our systems developed on English data generalise to Mandarin data.
We also evaluate the effect of an SSL model's pre-training language by testing with models trained on different amounts of Mandarin data.
Concretely, we compare the English-only WavLM Large~\cite{Chen2022}, the multilingual mHuBERT~\cite{Boito2024} including limited Mandarin data, and the Mandarin HuBERT Large trained entirely on Mandarin (see Section~\ref{sec:exp_setup_implementation} for details).
We show results for the $k$-means system representative of previous work~\cite{Malan2025}, together with the two systems that worked best on English (Table~\ref{tab:test_results}).

Across the different systems, the English-only WavLM performs worst, the mHuBERT being exposed to some Mandarin gives intermediate results, and the Mandarin HuBERT performs the best by a substantial margin.
In particular, Mandarin HuBERT features achieve a NED of 4.9\% when averaged and graph clustered.
This results in a lexicon of comparable quality to the best English lexicon (Table~\ref{tab:test_results}) 
and shows that including more of the target language in pre-training improves lexicon quality.
This underscores the importance of language-specific SSL representations~\cite{vanNiekerk2024, Malan2024}.

Although we show that quality lexicons can be obtained on a different language, the results here corroborate those on English where even the strongest systems produce lexicons with a NED of around 5\% and word-level purity of between 80\% and 90\%.
Below we examine the source of this limitation.

\section{Representations vs Clustering} 
\label{perfect_representations}

Given that we are using true word boundaries, lexicon imperfections stem 
from clustering methods that fail to group instances of the same word together and/or representations that fail to capture similarity across different instances of the same word.
In Table~\ref{tab:perf_representations_GT_numb} we present controlled experiments on LibriSpeech test-clean where we idealise either the clustering method or representations in two representative systems.
The baseline results are from Table~\ref{tab:test_results}.

\begin{table}[t!]
    \centering
    \caption{
    Lexicon quality (\%) on LibriSpeech test-clean when the effect of representations or clustering methods is isolated.
    }
    \tablesep
    \label{tab:perf_representations_GT_numb}
        \renewcommand{\arraystretch}{1.1}

    \begin{tabularx}{\linewidth}{@{}l@{\ \ }RRr@{}}
    \toprule
        System & {NED}  & {Purity} & {V-measure} \\
        \midrule
        \multicolumn{4}{c}{Continuous + average + $k$-means} \\
        \midrule
        Baseline & 8.6 & 88.4 & 83.6 \\
        Perfect cluster initialisation & 17.0 & 81.5 & 81.3 \\
        Perfect word embeddings & 12.1 & \textbf{100.0} & \textbf{100.0} \\
        \midrule
        \multicolumn{4}{c}{Discrete + edit + graph clustering} \\
        \midrule
        Baseline & 7.9 & 83.0 & 88.4 \\
        Perfect cluster initialisation & 7.4 & 83.6 & 88.7 \\
        Perfect word representations & 12.1 & \textbf{100.0} & \textbf{100.0} \\
        \bottomrule
        \end{tabularx}
\end{table}

In the second row of each section in the table, the two clustering algorithms are initialised so that all instances of the same word type are assigned to the same cluster.
For $k$-means (top section), this means that centroids are initialised at the mean embeddings of segments corresponding to the same word type.
For graph clustering (bottom), segments are initially assigned to clusters based on their true word types.
At initialisation, both systems have a purity and V-measure of 100\%.
However, after the methods are run from their perfect starting points, the resulting lexicons are far from ideal: purity and V-measure drop to baseline levels.
From the $k$-means system we conclude that embeddings of the same word type are not close to all the other instances of that word.
Similarly, the discrete system shows that unit sequences vary across instances of the same word type.

Next we consider the case where representations are ideal and clustering is performed as usual. 
For continuous features (top section), each word type is represented by its mean acoustic word embedding with a small amount of random noise added. 
For the discrete unit sequences (bottom), a single sequence is randomly selected to represent all instances of that word type. 
Results are given in the third row of each section in Table~\ref{tab:perf_representations_GT_numb}.
We see that, in both cases, a perfect lexicon is produced with word-level purities and V-measures of 100\%.
In both cases, NED is worse than the baselines, dropping from roughly 8\% to 12\%.
This is because phonetic variation, even in clusters consisting of only one word type, limits NED as a word-level metric. 
Conversely, word-level metrics, such as purity, ignore phonetic detail. This is why we report a range of complementary metrics.

Pronunciation and co-articulation not only affects NED but also lead to differences in the representations, which in part explains the poor performance when initialising the clustering methods perfectly (the second rows in the two sections).
But in the $k$-means systems we see that perfect initialisation actually leads to worse performance than the non-ideal baseline, e.g.\ purity dropping from 88.4\% to 81.5\%.
Together these findings suggest that clustering methods, particularly graph clustering, perform robustly, but inconsistency in the representation of spoken words hinder effective lexicon learning.
If consistent representations aligned with word types were available, perfect word-level purity and V-measure would be achievable.

\section{Conclusion}
\label{sec:conclusion}

This paper compared six systems for lexicon learning that combine different feature representations and clustering approaches, achieving clear improvements over previous methods.
These results summarise the current landscape of lexicon learning in terms of accuracy and computational cost.
In controlled experiments, we showed that if representations are idealised, existing
clustering methods can learn perfect lexicons.
The current bottleneck is therefore not in clustering, but in the consistency of feature representations.
Future work will explore how these insights can be applied to the training of self-supervised speech models to improve the resulting representations.
We will also consider how our best systems can be extended for true word discovery, where boundaries are~unknown.

\bibliography{refs}

\end{document}